\documentclass[final,3p,twocolumn]{elsarticle}

\usepackage{amssymb}
\usepackage{graphicx}
\usepackage{dcolumn}
\usepackage{bm}
\usepackage{longtable}
\usepackage{lineno,hyperref}

\usepackage[numbers]{natbib}

\begin{document}

\begin{frontmatter}

\title{De-Confinement in small systems: Clustering of color sources in high multiplicity $\bar{p}$p collisions at $\sqrt{s}$= 1.8 TeV }
%
\author[address3]{L. G. Gutay}
\author[address3]{A. S. Hirsch}
\author[address4]{C. Pajares}
\author[address3]{R. P. Scharenberg}
\author[address3]{B. K. Srivastava\corref{corrauthor}}
\cortext[corrauthor]{Corresponding author}
\ead{brijesh@purdue.edu}
\address[address3]{Department of Physics and Astronomy, Purdue University, West Lafayette, IN-47907, USA}
\address[address4]{Departamento de Fisica de Particulas, Universidale de Santiago de Compos
tela and Instituto Galego de Fisica de Atlas Enerxias(IGFAE), 15782 Santiago, de Compostela
, Spain}
\date{\today}
\begin{abstract}
It is shown that de-confinement can be achieved in high multiplicity non jet $\bar{p}$p collisions at $\sqrt{s}$= 1.8 TeV Fermi National Accelerator Laboratory(FNAL- E735) experiment. 
Previously the evidence for de-confinement was the demonstrated by the constant freeze out energy density in high multiplicity events. 
 In this paper we use the same data but analyze the transverse momentum spectrum in the framework of the clustering of color sources. The charged particle pseudorapidities  densities in the range 7.0 $\leq \langle dN_{c}/d\eta \rangle \leq$26.0 are considered. Results are presented for both thermodynamic and transport properties. The initial temperature and  energy density are obtained and compared with the Lattice Quantum Chromo Dynamics(LQCD) simulations. The energy density ($\varepsilon/T^{4}$) $\sim$ 11.5 for $ \langle dN_{c}/d\eta \rangle \sim $ 25.0 is close to the value for 0-10\% central events in Au+Au collisions at $\sqrt{s_{NN}}$= 200 GeV. The shear viscosity to entropy density ratio($\eta/s$) is $\sim$ 0.2 at the transition temperature.   
The result for the trace anomaly $\Delta$ is in excellent agreement with LQCD simulations. These results confirm our earlier observation that the de-confined state of matter was created in high multiplicity events in $\bar{p}$p collisions at  $\sqrt{s}$=1.8 TeV.  

\end{abstract}


\end{frontmatter}

\section{Introduction}

The observation of high total  multiplicity, high transverse energy, non-jet, isotropic events led Van Hove \cite{hove} and Bjorken \cite{boje} to conclude that high energy density events are produced in high energy $\bar{p}$p collisions \cite{larry}. These events have a far greater cross  section  than  the jet  production.  In  these  events the transverse energy is proportional to the number of low transverse momentum particles. This basic correspondence can be explored over a wide range of the charged particle pseudorapidity density  $ \langle dN_{c}/d\eta \rangle$ in $\bar{p}$p collisions at center of mass energy $\sqrt {s}$ = 1.8 TeV. 
The analysis of charged particle transverse momentum data from the E735 experiment exhibits flow velocity of mesons and anti-baryons also indicating the possible evidence of QGP formation \cite{muller}. 

Collective hydrodynamics flow has been successful in  explaining the $AA$ collisions at RHIC and LHC energies \cite{flow}. However $pp$ collisions have been considered different from the heavy ion collisions. 
The observation of long range rapidity correlations, the so called ``ridge'', similar to that seen in heavy ion collisions in high multiplicity $pp$ collisions at $\sqrt{s}$= 7 TeV by the CMS experiment suggests the evidence of strong radial flow \cite{CMS,shuryak}. There are several theoretical papers which support the view that QGP can be formed in high multiplicity $pp$ events \cite{werner,andres,ghosh}.

In our earlier work, published in 2002,  the evidence of hadronic de-confinement in $\bar{p}$p collisions at $\sqrt {s}$ = 1.8 TeV was presented \cite{e735d}. Based on the HBT analysis a constant freeze-out energy density $\sim$ 1 $GeV/fm^{3}$ for high multiplicity $\bar{p}$p events was measured \cite{e735d}. The freeze-out energy density was found to be independent of the $dN_{c}/d\eta$ $\geq 6 $.  

The objective of this work is to further analyze the published E735 data on the transverse momentum spectra of charged particles in the framework of  Color String Percolation Model(CSPM). CSPM has been successfully applied to heavy ion data for thermodynamics and transport coefficients \cite{eos,cpod13,eos2,IS2013,eos3}. 
After a brief description of the color string percolation model (CSPM)\cite{pajares1,pajares2} and the E735 experiment the results are presented for the shear viscosity to entropy density ratio($\eta/s$) and the Equation of State (EOS) of the deconfined matter. 

\section{String Interactions and Percolation}
Multiparticle production at high energies is currently described in terms of color strings stretched between the projectile and target. Hadronizing these strings produce the observed hadrons. The strings act as color sources of particles through the creation of $q \bar{q}$ pairs from the sea. At low energies only valence quarks of nucleons form strings that then hadronize. The number of strings grows with the energy and with the number of nucleons of participating nuclei. Color strings may be viewed as small discs in the transverse space filled with the color field created by colliding partons. Particles are produced by the Schwinger mechanisms \cite{swinger}.  With growing energy and size of the colliding nuclei the number of strings grow and start to overlap to form clusters \cite{pajares1,pajares2}. At a critical density a macroscopic cluster appears that marks the percolation phase transition. 2D percolation is a non-thermal second order phase transition. In CSPM the Schwinger barrier penetration mechanism for particle production and the fluctuations in the associated string tension due to the strong string interactions make it possible to define a temperature.
Consequently the particle spectrum is "born" with a thermal distribution \cite{bialas}. With an increasing number of strings there is a progression from isolated individual strings to clusters and then to a large cluster which suddenly spans the area. In two dimensional percolation theory the relevant quantity is the dimensionless percolation density parameter given by \cite{pajares1,pajares2}  
\begin{equation}  
\xi = \frac {N S_{1}}{S_{\bot}}
\end{equation}
where N is the number of strings formed in the collisions and $S_{1}$ is the transverse area of a single string and $S_{\bot}$ is the transverse nuclear overlap area. The critical cluster which spans $S_{\bot}$, appears for
$\xi_{c} \ge$ 1.2 \cite{isich,satz1}. As $\xi$ increases the fraction of $S_{\bot}$ covered by this spanning cluster increases.

We assume that a cluster of $n$ strings behaves as a single string with an energy-momentum that corresponds to the sum of energy-momenta of the individual strings and with a higher color field, corresponding to the vectorial sum of the color field of each individual string \cite{pajares1,pajares2}. One can obtain the multiplicity $\mu$ and the mean transverse momentum squared $\langle p_{t}^{2} \rangle$ of the particles produced by a cluster of $\it n $ strings \cite{pajares2}  
\begin{equation}
\mu_{n} = \sqrt {\frac {n S_{n}}{S_{1}}}\mu_{0};\hspace{5mm}
\langle p_{t}^{2} \rangle = \sqrt {\frac {n S_{1}}{S_{n}}} {\langle p_{t}^{2} \rangle_{1}}
\end{equation} 
where $\mu_{0}$ and $\langle p_{t}^{2}\rangle_{1}$ are the mean multiplicity and $\langle p_{t}^{2} \rangle$ of particles produced from a single string with a transverse area $S_{1} = \pi r_{0}^2$. In the limit of high string density, one obtains \cite{pajares1,pajares2}
\begin{equation}
\langle \frac {n S_{1}}{S_{n}} \rangle = \frac {\xi}{1-e^{-\xi}}\equiv \frac {1}{F(\xi)^2}
\end{equation}

where $F(\xi)$ is the color suppression factor.  $F(\xi)$ is related to the $\xi$.
\begin{equation}
F(\xi) = \sqrt {\frac {1-e^{-\xi}}{\xi}.}
\end{equation}
The net effect due to $F(\xi)$ is the reduction in hadron multiplicity and increase in the average transverse momentum of particles. The CSPM model calculation for hadron multiplicities and momentum spectra were found to be in excellent agreement with experiment \cite{IS2013}.

It is worth noting that CSPM is a saturation model similar to the Color Glass Condensate (CGC),  where $ {\langle p_{t}^{2} \rangle_{1}}/F(\xi)$ plays the same role as the saturation momentum scale $Q_{s}^{2}$ in the CGC model \cite{cgc,perx}. 

\section{ E735 experiment }
  Experiment E735 was run during the 1988-1989 Tevatron running period, primarily collecting data triggered to enrich high multiplicity events in  $\bar{p}$ collisions at  $\sqrt{s}$ = 1.8 TeV \cite{e735a}. The E735 was located at the $C_{\phi}$ interaction region of the Fermi National Accelerator Laboratory (FNAL) \cite{e735a,e735b,e735c}. The $\bar{p}$p interaction region was surrounded by a cylindrical drift chamber which in
turn was covered by a single layer hodoscope including endcaps. This system measured the total charged particle multiplicity 10 $ < N_{c} < $ 200 in the pseudorapidity range $|\eta| < $3.25. A sidearm magnetic spectrometer with tracking chambers and time of flight counters, provided particle identified momenta spectra in the
range 0.1 $< p_{t} < $ 1.5 GeV/c. The spectrometer covered −0.37 $ <\eta <$ +1.00 with $\Delta \phi \sim∼ 20^{0}$ ($\phi$ is the azimuthal angle around the beam direction).

 The multiplicity dependence of the transverse momentum $p_{t}$ spectra was measured in $\bar{p}$p collisions at $\sqrt{s}$ = 1.8 TeV\cite{e735c}. The invariant $p_{t}$ spectra was fitted with a power law $A/(p_{0}+p_{t})^{n}$ \cite{ua1}. Table \ref{tab1} shows the power law fit parameters for high multiplicity events \cite{e735c}. 
\begin{table}[thbp]
\caption{Number of tracks $N_{c}$ as measured by the E735 experiment in the pseudorapidity range $|\eta| < $3.25, $\langle dN_{c}/d\eta \rangle $, and the fit parameters $p_{0}$ and ${\it n}$ to the invariant $p_{t}$ distribution \cite{e735b,e735c}.}
\vspace*{0.6cm}
\setlength{\tabcolsep}{4pt}
\begin{tabular}{|c|c|c|c|c|}\hline
$N_{c}$ & $\langle dN_{c}/d\eta \rangle $ & $ p_{0}$  & ${\it n}$ \\ \hline
 47 (minbias) & 7.3   & 1.25   & 8.35  \\ \hline
 85           & 13.07 & 1.052  & 7.038  \\ \hline
 105          & 16.15 & 1.001  & 6.743   \\ \hline
 135          & 20.76 & 1.001  & 6.581    \\ \hline
 165          & 25.38 & 1.061  & 6.766     \\ \hline
\end{tabular}
\label{tab1}
\end{table}

\section{ Determination of  the color suppression factor F($\xi$) }
 The suppression factor is determined by comparing the charged particle spectra from low energy ${\it pp}$ collisions and  high multiplicity  transverse momentum spectra from $\bar{p}$p. To evaluate the initial value of $\xi$ from data  a parameterization of ${\it pp}$ events at $\sqrt{s}$ = 200 GeV  is also used to compute the $p_{t}$ distribution 
\begin{equation}
dN_{c}/dp_{t}^{2} = a/(p_{0}+p_{t})^{\it {n}}
\end{equation}
where a is the normalization factor. $p_{0}$ = 1.98 and ${\it n}$ = 12.88 are parameters used to fit the data. This parameterization is used for high multiplicity events in $\bar{p}$p collisions at $\sqrt{s}$ = 1.8 TeV to take into account the interactions of the strings \cite{pajares2}.
\begin{equation}
dN_{c}/dp_{t}^{2} = \frac {a'}{{(p_{0}{\sqrt {F(\xi_{pp})/F(\xi_{\bar{p}p})}}+p_{t})}^{\it {n}}}
\end{equation}

\begin{figure}[thbp]
\centering        
\vspace*{-0.2cm}
\includegraphics[width=0.50\textwidth,height=3.0in]{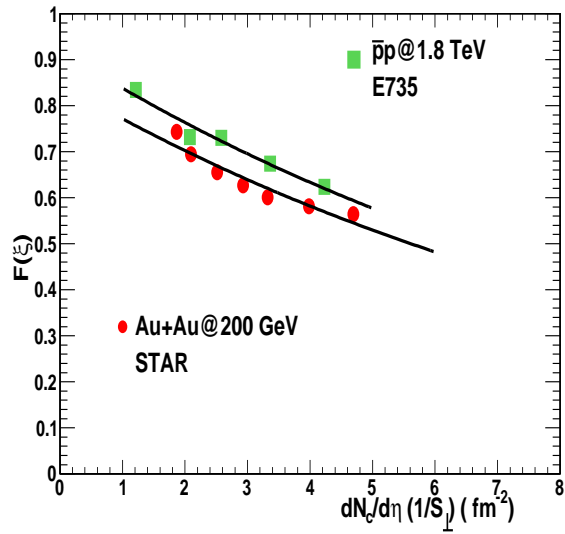}
\vspace*{-0.5cm}
\caption{The color suppression factor $F(\xi)$ as a function of $\frac {dN_{c}}{d\eta}/S_{\bot}(fm^{-2})$. The solid green squares are from $\bar{p}$p at $\sqrt {s}$=1.8 TeV. 
The solid red circles are for Au+Au collisions at $\sqrt {s_{NN}}$ = 200 GeV(STAR data) \cite{eos2}. The error is smaller than the size of the symbol. Lines are fit to $\bar{p}$p and the STAR data.}
\label{suppression}
\end{figure}

In pp collisions $F(\xi_{pp}) \sim$ 1 at $\sqrt {s}$ = 200 GeV due to the low overlap probability \cite{eos}. Figure (\ref{suppression}) shows a plot of $F(\xi)$ as a function of charged particle multiplicity per unit transverse area $\frac {dN_{c}}{d\eta}/S_{\bot}$ for high multiplicity events along with the minbias events.
For $pp$ collisions the overlap area $S_{\bot}$ is taken as the inelastic $pp$ cross section $\sigma_{pp}$ \cite{levente,cdf,alice}. For $\bar{p}$p collision at $\sqrt{s}$=1.8 $\sigma_{\bar{p}p} \sim $ 60 mb \cite{cdf}. 
The error on  $F(\xi)$ is $\sim 6\%$.
The transverse overlap area measured in E735 by Hanbury-Brown-Twiss is in excellent agreement with this inelastic cross section \cite{e735d}. 
The results from Au+Au collisions at $\sqrt s_{NN}$ = 200 GeV \cite{eos2,eos3} are also shown in Fig. (\ref{suppression}). 
\section{Temperature}
The connection between the measured $F(\xi)$ and the temperature $T(\xi)$ involves the Schwinger mechanism (SM) for particle production. 
The Schwinger distribution for massless particles is expressed in terms of $p_{t}^{2}$ \cite{swinger,wong}
\begin{equation}
dn/d{p_{t}^{2}} \sim e^{-\pi p_{t}^{2}/x^{2}}
\end{equation}
where the average value of the string tension is  $\langle x^{2} \rangle$. The tension of the macroscopic cluster fluctuates around its mean value because the chromo-electric field is not constant.
The origin of the color string fluctuation is related to the stochastic picture of 
the QCD vacuum. Since the average value of the color field strength must 
vanish, it can not be constant but changes randomly from point to point \cite{bialas}. Such fluctuations lead to a Gaussian distribution of the string tension for the cluster, which gives rise to the thermal distribution \cite{eos,bialas,pajares3}
\begin{equation}
dn/d{p_{t}^{2}} \sim e^{(-p_{t} \sqrt {\frac {2\pi}{\langle x^{2} \rangle}} )}
\end{equation}
with $\langle x^{2} \rangle$ = $\pi \langle p_{t}^{2} \rangle_{1}/F(\xi)$. 
The temperature is expressed as  
\begin{equation}
T(\xi) =  {\sqrt {\frac {\langle p_{t}^{2}\rangle_{1}}{ 2 F(\xi)}}}
\label{temp}
\end{equation} 
where $\langle p_{t}^{2}\rangle_{1}$  is the average transverse momentum squared of particles produced from a single string. At the percolation transition $\xi$ = 1.2 and the critical value of the temperature is taken $T_{c}$ = 167 MeV from the statistical model, which is the universal chemical freeze-out temperature and is a good measure of the phase transition temperature \cite{fr1,fr2}. This determines $\sqrt {\langle p_{t}^{2}\rangle_{1}}$ = 207 MeV \cite{eos}.
The temperature obtained using Eq. (\ref{temp}) for $\langle dN_{c}/d\eta \rangle \sim $ 25 is $\sim$ 188 MeV. This temperature is closed to  $\sim$ 193.6 MeV obtained for 0-10 \%Au+Au collisions at $\sqrt s_{NN}$ = 200 GeV\cite{eos}. 

\begin{table}
\caption{$\langle dN_{c}/d\eta \rangle $,  the measured percolation density parameter $\xi$, initial temperature $T$, initial energy density $\varepsilon$  and $\eta/s$ for $\bar{p}$p $\sqrt {s}$=1.8 TeV}
\vspace*{0.5cm}
\setlength{\tabcolsep}{4pt}
\begin{tabular}{|c|c|c|c|c|}\hline
$\langle dN_{c}/d\eta \rangle $ & $\xi$  & T (MeV) & $\varepsilon (GeV/fm{^3})$ & $\eta/s$ \\ \hline
 7.30(minbias) & 0.778 & 160.31 & 0.50 & 0.30  \\ \hline
 13.07 & 1.39  &  170.73 & 0.86 & 0.23   \\ \hline
 16.15 & 1.42  & 171.22 &  1.07 & 0.23    \\ \hline
 20.76 & 1.84 &  178.06 & 1.39 & 0.21    \\ \hline
 25.38 & 2.30 &  185.07 & 1.75 & 0.21      \\ \hline
\end{tabular}
\label{tab2}
\end{table}
\section{Energy Density }
Among the most important and fundamental problems in finite-temperature QCD are the calculation of the bulk properties of hot QCD matter and characterization of the nature of the QCD phase transition. 
The QGP according to CSPM is born in local thermal equilibrium  because the temperature is determined at the string level. After the initial temperature $ T > T_{c}$ the  CSPM perfect fluid may expand according to Bjorken boost invariant 1D hydrodynamics \cite{bjorken}
\begin{equation}
\varepsilon = \frac {3}{2}\frac { {\frac {dN_{c}}{dy}}\langle m_{t}\rangle}{S_{\bot} \tau_{pro}}
\end{equation}
where $\varepsilon$ is the energy density, $S_{\bot}$ the transverse overlap area of the colliding nuclei, and $\tau_{pro}$ the proper time. Above the critical temperature only massless particles are present in CSPM. To evaluate $\varepsilon$ we use the charged pion multiplicity $dN_{c}/{dy}$. The factor 3/2 in Eq.(12) accounts for the neutral pions. The average transverse mass $\langle m_{t}\rangle$ is given by $\langle m_{t}\rangle =\sqrt {\langle p_{t}\rangle^2 + m_{0}^2} $, where $\langle p_{t}\rangle$ is the average transverse momentum of pion and $m_{0}$ the mass of pion. In Schwinger model $\tau_{pro}$ is given by $\tau_{pro} = \frac {2.405\hbar}{\langle m_{t}\rangle}$ \cite{swinger}.

 The pion multiplicity was obtained from the total multiplicity after subtracting the contributions due to kaons and protons \cite{e735a,e735}. The average $p_{t}$ $\langle p_{t}\rangle$ for pions are flat in the region 7 $\leq \langle dN_{c}/d\eta \rangle \leq $ 26. Table \ref{tab2} shows the values of $\varepsilon$ and temperature for various values of $\langle dN_{c}/d\eta \rangle$. 
\begin{figure}[thbp]
\centering        
\vspace*{-0.2cm}
\includegraphics[width=0.50\textwidth,height=3.0in]{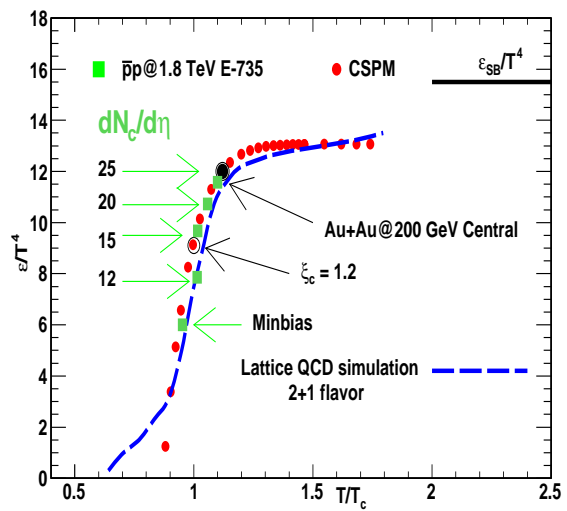}
\vspace*{-0.5cm}
\caption{$\varepsilon/T^{4}$ as a function of T/$T_{c}$. Solid green squares are from $\bar{p}$p collisions at $\sqrt{s}$ = 1.8 TeV. The lattice QCD calculation is shown as dotted blue line \cite{hotqcd}. CSPM for Au+Au at  $\sqrt {s_{NN}}$ = 200 GeV are red solid circles \cite{eos2}.}
\label{et4}
\end{figure}
 With the determination of $T$ and  $\varepsilon$ one can  compare the energy density expressed as $\varepsilon/T^{4}$ with the available lattice QCD results 
\cite{hotqcd}. Figure \ref{et4} shows a plot of $\varepsilon/T^{4}$  as a function of T/$T_{c}$. 
The lattice QCD results are from the HotQCD Collaboration 
\cite{hotqcd}.  The result for most central collisions for Au+Au at $\sqrt {s_{NN}}$ = 200 GeV is also shown in Fig.\ref{et4}. It is observed that for highest multiplicity $\langle dN_{c}/d\eta \rangle \sim$ 27,  $\varepsilon/T^{4}$ is close to the value obtained for Au+Au at 200 GeV. The minimum bias result is below the percolation threshold. This result confirms the formation of the QGP in high multiplicity events in $\bar{p}$p interactions at $\sqrt{s}$ = 1.8 TeV.   

\section{ Shear viscosity to entropy density ratio $\eta$/s  }
In our earlier work the shear viscosity to entropy density ratio $\eta/s$ was obtained in the framework of kinetic theory and the string percolation \cite{eos2}. The following expression was  obtained for $\eta/s$ \cite{gul2,gul1,eos2}. 
\begin{equation}
\frac {\eta}{s} ={\frac {TL}{5(1-e^{-\xi})}} 
\label{visco}
\end{equation}
where T is the temperature and L is the longitudinal extension of the source $\sim$ 1 fm.

\begin{figure}[thbp]
\centering        
\vspace*{-0.2cm}
\includegraphics[width=0.50\textwidth,height=3.0in]{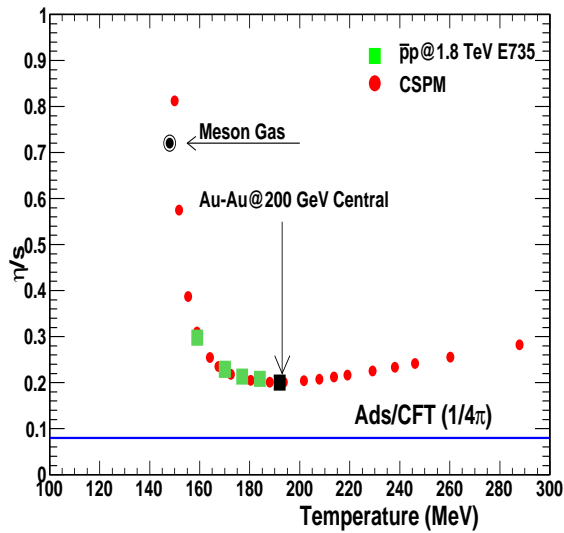}
\vspace*{-0.5cm}
\caption{$\eta/s$ as a function of temperature T using Eq. (\ref{visco}) \cite{eos2}. Solid green squares are from $\bar{p}$p collisions at $\sqrt{s}$ = 1.8 TeV. CSPM results for Au+Au at $\sqrt {s_{NN}}$ = 200 GeV are shown as solid red circles \cite{eos2}. The meson gas value for $\eta/s$ $\sim$ 0.7 is shown as solid black circle at T $\sim$ 150 MeV \cite{prakash}. The lower bound shown is given by the AdS/CFT \cite{kss}.} 
\label{vis}
\end{figure}

Fig.~\ref{vis} shows $\eta/s$ as a function of the temperature \cite{eos2}. The lower bound shown in Fig.~\ref{vis} is given by the AdS/CFT conjecture \cite{kss}. 

The results from Au+Au at $\sqrt {s_{NN}}$ = 200 GeV collisions are also shown in Fig.~\ref{vis}  for comparison purposes. It is seen that for $\langle dN_{c}/d\eta \rangle \sim$ 25 the $\eta/s$ is equal to the value from the most central collisions in Au+Au at 200 GeV.   
\section{ $\eta/s$ and Trace anomaly $\Delta$}
\begin{figure}[thbp]
\centering        
\vspace*{-0.2cm}
\includegraphics[width=0.50\textwidth,height=3.0in]{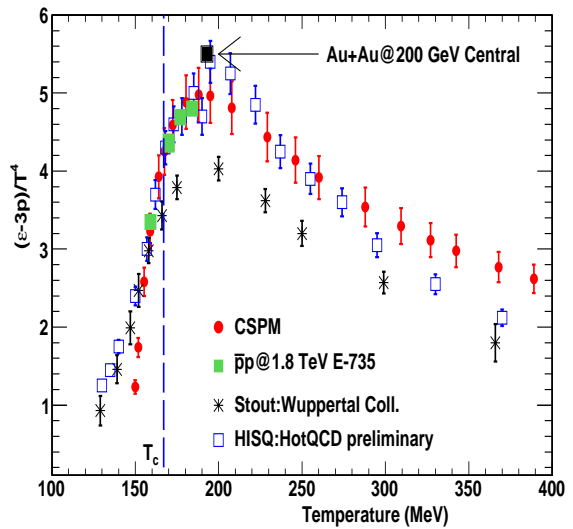}
\vspace*{-0.5cm}
\caption{The trace anomaly $\Delta =(\varepsilon-3p)/T^{4}$ vs temperature \cite{lattice12}. Solid green squares are from $\bar{p}$p collisions at $\sqrt{s}$ = 1.8 TeV. Blue open squares are from HotQCD Collaboration \cite{lattice12}. Black stars are from Wuppertal Collaboration \cite{wuppe}.} 
\label{trace}
\end{figure}

The trace anomaly ($\Delta$) is the expectation value of the trace of the energy-momentum tensor, $\langle \Theta_{\mu}^{\mu}\rangle = (\varepsilon-3p)$, which measures the deviation from conformal behavior and thus identifies the interaction still present in the medium \cite{cheng}. We find that the reciprocal of $\eta/s$ is in quantitative agreement with $(\varepsilon-3p)/T^{4}$ from LQCD over a wide range of temperatures \cite{cpod13,IS2013}.  Fig.~\ref{trace} shows the $\Delta$ for $\bar{p}$p along with the CSPM calculation and LQCD simulations. The minimum in $\eta/s \sim 0.20$ determines the peak of the interaction measure $\sim$ 5 in agreement with the recent HotQCD values \cite{lattice12}. Figure~\ref{trace} also shows the results from the Wuppertal Collaboration \cite{wuppe}. 
The maximum in $\Delta$ corresponds to the minimum in $\eta/s$. This is true for the highest multiplicity events in $\bar{p}$p collisions.
\section{Equation of State: Sound velocity}
We use CSPM coupled to a 1D Bjorken expansion. The input parameters are the initial temperature T, the initial energy density $\varepsilon$, and the trace anomaly $\Delta$ are determined by data. The Bjorken 1D expansion far the sound velocity can be written \cite{bjorken}
\begin{equation}
\frac {dT}{d\varepsilon} s = C_{s}^{2}. 
\label{sound}
\end{equation}
In the above equation the entropy density $s$ is expressed as $s = (\varepsilon + P)/T$. The pressure $P$ is related to trace anomaly $P = (\varepsilon-\Delta T^{4})/3$. We can express $C_{s}^{2}$ in terms of $\xi$ 

\[
 C_{s}^{2} = (-0.33)\left(\frac {\xi e^{-\xi}}{1- e^{-\xi}}-1\right)\]\begin{equation}
 + 0.0191(\Delta/3)\left(\frac {\xi e^{-\xi}}{({1- e^{-\xi}})^2}-\frac {1}{1-e^{-\xi}} \right)
\label{sound2}
\end{equation}
\begin{figure}[thbp]
\centering        
\vspace*{-0.2cm}
\includegraphics[width=0.50\textwidth,height=3.0in]{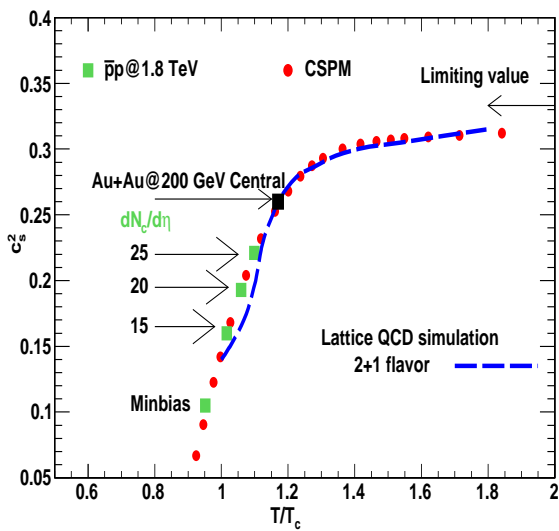}
\vspace*{-0.5cm}
\caption{The speed of sound from CSPM (red circles) and Lattice QCD-p4 (blue dash line)  versus $T/T_{c}$\cite{hotqcd}. The green solid squares are from $\bar{p}$p at $\sqrt{s}$ = 1.8 TeV.}
\label{cs2}
\end{figure}
Since there is no direct way to obtain pressure in CSPM, we have made the assumption that the dimensionless ratio $\Delta = (\varepsilon- 3 P)/T^{4} \approx 1/(\eta/s)$. Fig.~\ref{cs2} shows a plot of $C_{s}^{2}$ as a function of $T/T_{c}$. It is observed that the CSPM results are in very good agreement with the lattice calculations \cite{hotqcd}. This suggests that the $\Delta$ as obtained from the data can be represented by $1/(\eta/s)$.
\section{Summary}

The $\bar{p}$p collisions at $\sqrt{s}$ = 1.8 TeV from the E735 experiment have been analyzed in using the clustering of color sources phenomenology. The results for temperature, energy density and transport coefficient have been extracted for high multiplicity events. The trace anomaly is also obtained and compared with the LQCD results.  The clustering of color sources has shown that the determination of $\eta/s$ as a function of temperature is an important quantity that relates to the trace anomaly $\Delta$. The main assumption of the present approach is  that the inverse of $\eta/s$ represents the trace anomaly,  $\Delta = (\varepsilon -3{\it p)/T^{4}}$. The clustering of color sources (percolation) provides us with a microscopic partonic picture that also connects the transport properties of the QGP to its thermodynamics. 
 
 These results strongly suggest that even in small systems at high energy and high multiplicity events QGP formation is possible as seen in $\bar{p}p$ collisions at $\sqrt{s}$ = 1.8 TeV. 

The analysis of the  $pp$ collisions at $\sqrt{s}$ = 7 and 14 TeV at the CERN LHC can map event higher temperatures and energy densities. Then it will be possible to obtain the trace anomaly at higher temperature as shown in Fig.(\ref{trace}). The attenuation of high $p_{t}$ particles in high multiplicity events would be a second signal that the de-confined matter is indeed the QGP.  

\section{Acknowledgment}
This research was supported by the Office of Nuclear Physics within the U.S. Department of Energy  Office of Science under Grant No. DE-FG02-88ER40412. CP thanks Ministry of Economy of Spain and FEDER funds for financial support.
%

\bibliography{all}

\begin{thebibliography}{10}
\expandafter\ifx\csname url\endcsname\relax
  \def\url#1{\texttt{#1}}\fi
\expandafter\ifx\csname urlprefix\endcsname\relax\def\urlprefix{URL }\fi
\expandafter\ifx\csname href\endcsname\relax
  \def\href#1#2{#2} \def\path#1{#1}\fi

\bibitem{hove}
L.~V. Hove, Phys. Lett. B 118 (1982) 138.

\bibitem{boje}
J.~D. Bjorken, Fermilab Pub. 82/44-THY.

\bibitem{larry}
L.~McLerran, Rev. Mod. Phys. 58 (1986) 1021.

\bibitem{muller}
P.~Levai, B.~Muller, Phys. Rev. Lett. 67 (1991) 1519.

\bibitem{flow}
U.~Heinz, R.~Snellings, Ann. Rev. Nucl. Part. Sci 63 (2013) 123.

\bibitem{CMS}
V.~{Khachatryan et al. (CMS Collaboration)}, JHEP 1009 (2010) 091.

\bibitem{shuryak}
T.~Kalaydzhyan, E.~Shuryak, arXiv:1503.0521.

\bibitem{werner}
K.~Werner, B.~Guiot, I.~Karpenko, T.~Piergo, Nucl. Phys. A 931 (2014) 83.

\bibitem{andres}
C.~Andres, A.~Moscoso, C.~Pajares, Phys. Rev. C 90 (2014) 054902.

\bibitem{ghosh}
P.~Ghosh, S.~Muhuri, J.~K. Nayak, R.~Varma, J. Phys. G 41 (2014) 03516.

\bibitem{e735d}
T.~{Alexopoulos et al. (E735 Collaboration)}, Phys. Lett. B 528 (2002) 43.

\bibitem{eos}
R.~P. Scharenberg, B.~K. Srivastava, A.~S. Hirsch, Eur. Phys. J. C 71 (2011)
  1510.

\bibitem{cpod13}
R.~P. Scharenberg, PoS (CPOD 2013) 017.

\bibitem{eos2}
J.~D. de~Deus, A.~S. Hirsch, C.~Pajares, R.~P. Scharenberg, B.~K. Srivastava,
  Eur. Phys. J. C 72 (2012) 2123.

\bibitem{IS2013}
B.~K. Srivastava, {Nucl. Phys. A} 926 (2014) 142.

\bibitem{eos3}
M.~A. Braun, J.~D. de~Deus, A.~S. Hirsch, C.~Pajares, R.~P. Scharenberg, B.~K.
  Srivastava, arXiv:1501.01524.

\bibitem{pajares1}
M.~A. Braun, C.~Pajares, Eur. Phys. J. C 16 (2000) 349.

\bibitem{pajares2}
M.~A. Braun, F.~del Moral, C.~Pajares, Phys. Rev. C 65 (2002) 024907.

\bibitem{swinger}
J.~Schwinger, Phys. Rev. 128 (1962) 2425.

\bibitem{bialas}
A.~Bialas, Phys. Lett. B 466 (1999) 301.

\bibitem{isich}
M.~B. Isichenko, Rev. Mod. Phys. 64 (1992) 961.

\bibitem{satz1}
H.~Satz, Rep. Prog. Phys. 63 (2000) 1511.

\bibitem{cgc}
L.~McLerran, R.~Venugopalan, Phys. Rev. D 49 (1994) 2233.

\bibitem{perx}
J.~D. de~Deus, C.~Pajares, Phys. Lett. B 695 (2011) 211.

\bibitem{e735a}
T.~{Alexopoulos et al. (E735 Collaboration)}, Phys. Rev. D 48 (1993) 984.

\bibitem{e735b}
T.~{Alexopoulos et al. (E735 Collaboration)}, Phys. Lett. B 336 (1994) 599.

\bibitem{e735c}
T.~{Alexopoulos et al. (E735 Collaboration)}, Phys. Lett. B 435 (1998) 453.

\bibitem{ua1}
C.~{Albajar et al. (UA1 Collaboration)}, Nucl. Phys. B 335 (1990) 261.

\bibitem{levente}
B.~I. {Abelev et al. (STAR Collaboration)}, Phys. Rev. C 79 (2009) 34909.

\bibitem{cdf}
S.~Klimenko, J.~Konigsberg, T.~M. Liss, Fermilab-FN-0741(2003).

\bibitem{alice}
B.~{Abelev et al. (ALICE Collaboration)}, Eur. Phys. J. C 73 (2013) 2456.

\bibitem{wong}
C.~Y. Wong, Introduction to high energy heavy ion collisions, 1994.

\bibitem{pajares3}
J.~D. de~Deus, C.~Pajares, Phys. Lett. 642 (2006) 455.

\bibitem{fr1}
P.~Braun-Munzinger, J.~Stachel, C.~Wetterich, Phys. Lett. B 596 (2004) 61.

\bibitem{fr2}
F.~Becattini, P.~Castorina, A.~Milov, H.~Satz, Eur. Phys. J. C 66 (2010) 377.

\bibitem{bjorken}
J.~D. Bjorken, Phys. Rev. D 27 (1983) 140.

\bibitem{e735}
T.~{Alexopoulos et al. (E735 Collaboration)}, Phys. Rev. Lett. 64 (1990) 991.

\bibitem{hotqcd}
A.~{Bazavov et al.}, Phys. Rev. D 80 (2009) 014504.

\bibitem{gul2}
P.~Danielewicz, M.~Gyulassy, Phys. Rev. D 31 (1985) 53.

\bibitem{gul1}
T.~Hirano, M.~Gyulassy, Nucl. Phys. A 769 (2006) 71.

\bibitem{kss}
P.~K. Kovtun, D.~T. Son, A.~O. Starinets, Phys. Rev. Lett. 94 (2005) 111601.

\bibitem{prakash}
M.~Prakash, M.~Prakash, R.~Venugopalan, G.~Wleke, Phys. Rep. 227 (1993) 321.

\bibitem{cheng}
M.~{Cheng et al.}, Phys. Rev. D 80 (2010) 054504.

\bibitem{lattice12}
P.~Petreczky, Lattice 2012 Cairns, Australia~(24 - 30 June).

\bibitem{wuppe}
S.~{Borsanyi et al.}, Phys. Rev. D 80 (2009) 014504.

\end{thebibliography}

\end{document}